\documentclass{article}
\usepackage{hiph-art}
\usepackage{graphicx}
\usepackage{amssymb}
\volnumber{19} \issuenumber{1} \edyear{2004}                             
\frompage{000} \topage{000}                                              
\recrevdate{1 January 2004}                                              
\def\rightmark{$T-\mu$ phase diagram of the chiral quark model}
\def\leftmark{A. Jakov\'ac, A. Patk{\'o}s, Zs. Sz{\'e}p, P. Sz{\'e}pfalusy}

\def\be{\begin{equation}}
\def\ee{\end{equation}}
\newcommand{\bea}{\begin{eqnarray}}
\newcommand{\eea}{\end{eqnarray}}

\title{Analytic determination of the $T-\mu$ phase diagram 
of the chiral quark model} 
\authors{ 
{A. Jakov\'ac$^{1,a}$, A. Patk{\'o}s$^{2,b}$, Zs. Sz{\'e}p$^{3,c}$ and
P. Sz{\'e}pfalusy$^{4,d}$ %
\index{A. Jakov\'ac} 
\index{A. Patk{\'o}s} 
\index{Zs. Sz{\'e}p}
\index{P. Sz{\'e}pfalusy}
}\\[2.812mm]
{\normalsize
\hspace*{-8pt}$^1$ Hungarian Academy of Sciences and Budapest University of
Technology and Economics, Research Group ``Theory of Condensed Matter'',
\\H-1521 Budapest, Hungary
\\[0.2ex] 
\hspace*{-8pt}$^2$ Department of Atomic Physics, E{\"o}tv{\"o}s University,\\
H-1117 Budapest, Hungary
\\[0.2ex]
\hspace*{-8pt}$^3$ Research Group for Statistical Physics of the
Hungarian Academy of Sciences,
\\H-1117 Budapest, Hungary
\\[0.2ex]
\hspace*{-8pt}$^4$ Department of Physics of Complex Systems,
E{\"o}tv{\"o}s University, \\H-1117 Budapest, Hungary\\
Research Institute for Solid State Physics and Optics,
Hungarian Academy of Sciences, H-1525 Budapest, Hungary
}}
 
\abstract{Using a gap equation for the pion mass a nonperturbative method is
given for solving  the chiral quark-meson model in the chiral limit at the
lowest order in the fermion contributions encountered in a large
N$_f$ approximation. The location of the tricritical point is
analytically determined. A mean field potential is constructed from which
critical exponents can be obtained.}

\keyword{large N approximation, tricritical point, finite density}

\PACS{11.10.Wx, 11.30.Rd, 25.75.Nq
}

\makeindex
\begin{document}
 
\maketitle

\section{Introduction}\label{sec:intro}

Due to the possibility of exploring parts of the phase diagram of strongly
interacting matter in current (RHIC) or forthcoming (LHC, GSI)
experiments, both numerical (lattice) and analytical investigations have
recently received much attention. Theoretically, the phase boundary of QCD
depends on the number of quark flavors taken into account and the masses of
the quarks \cite{QCD_map}. At vanishing chemical potential 
QCD with two massless quarks undergoes as a function of temperature a second order transition
from a hadron to a quark-gluon dominated phase.  At zero temperature and
large values of the chemical potential the transition from hadronic to more
exotic phases is of first order. With increasing temperature, the line of
the first order transitions ends in a tricritical point (TCP). For
non-vanishing masses of the u and d quarks TCP becomes a critical end point
(CEP) which separates the first order line from a crossover line
defined by the locations of finite maxima of some susceptibilities. The
decreasing value of the strange quark has the same effect as increasing
the value of the chemical potential, namely it turns the transition into a
first order one. Therefore, one can expect that for a physical value of
the strange quark mass the CEP in the $\mu-T$ plane will be closer to the 
temperature axis.

To locate TCP/CEP nonperturbative methods are required. First principle
lattice investigations using MC methods based on importance sampling are
made useless by the fact that the fermion determinant in the partition function
is complex at non-vanishing baryonic chemical potential. To avoid this
problem, recently new techniques were developed (for a review see
  e.g. \cite{LQCD_rev}) to extract, in the
$\mu/T\lesssim 1$ region, information for finite $\mu$ from simulations at zero
chemical potential (multiparameter reweighting of observables, Taylor
expansion of the fermion determinant) or at imaginary chemical potential,
followed by analytical continuation with or without Taylor expansion. Using
the multiparameter reweighting method the CEP was obtained for
$n_f=2+1$ dynamical staggered quarks with physical masses in \cite{katz04}.
The location of CEP was estimated in the same range in \cite{Ejiri_nara}
using an extrapolation based on simulation performed with 3 degenerate
quarks.

Analytically, the phase structure of QCD and the location of TCP/CEP can be
obtained using the Schwinger-Dyson approach \cite{QCD_SD} or the strong
coupling expansion of the lattice action \cite{strongQCD}. Alternatively, one
can investigate effective models of QCD \cite{LEeff}. Due to its low energy
nature, an effective model might prove not satisfactory in giving the exact
location of the endpoint and the shape of the phase boundary, but one can
still expect to obtain some insight into the physics near TCP/CEP such as
nonequilibrium dynamics. It can tell us what is the soft mode at the CEP,
and by predicting specific signatures \cite{sign} it may help finding its
location. Traces of a possible overlap between the critical regions of
TCP  and  CEP manifesting itself in the $\mu-T$ plane as
discussed in \cite{Ikeda03} could be in principle verified experimentally 
since it would appear as a gradual change in the value of the critical
exponents as the systems pass closer and closer to CEP. 

The outline of this paper is as follows. In sec. \ref{sec:model} we introduce 
an effective quark-meson model and present our method for its solution. 
In sec. \ref{sec:analysys}
we summarise the main results obtained for the phase diagram and 
investigate the relation between the effective potential of the theory
and the conventional treatment of the TCP in terms of a mean field 
Landau-type potential. We conclude in sec. \ref{sec:concl}.

\section{The model and the large $N_f$ method of its solution}\label{sec:model}

In the broken symmetry phase characterised by the homogeneous vacuum
expectation value $\Phi$ the usual shift in the radial direction
$\sigma \rightarrow \sqrt{N}\Phi+\sigma$
and the requirement of a finite constituent quark mass $m_q=g\Phi$ as 
$N\rightarrow \infty$ defines the following effective 
$SU_L(N_f)\times SU_R(N_f)$ symmetric Lagrangian for $N$ mesonic and $N_f$
fermionic fields ($N=N_f^2-1$):  
\bea
L[\sigma,\pi^a,\psi]=
-\left[\frac{\lambda}{24}\Phi^4+\frac{1}{2}m^2\Phi^2\right]N-
\left[\frac{\lambda}{6}\Phi^3+m^2\Phi\right]\sigma\sqrt{N}
\nonumber\\
+\frac{1}{2}\bigl[(\partial\sigma)^2 + (\partial\vec\pi)^2 \bigr]
-\frac{1}{2}m^2_{\sigma 0}\sigma^2
-\frac{1}{2}m^2_{\pi 0}\vec\pi^2
-\frac{\lambda}{6\sqrt{N}}\Phi\sigma \rho^2-
\frac{\lambda}{24N}\rho^4 \nonumber\\
+\bar \psi\left[i\partial^\mu\gamma_\mu-m_q-\frac{g}{\sqrt{N}}
\left(\sigma
+i\sqrt{2N_f}\gamma_5T^a\pi^a\right)\right]\psi+
\delta L_{ct}[\sigma,\pi^a,\psi],
\eea
where 
$m^2_{\sigma 0}=m^2+\frac{\lambda}{2}\Phi^2$,
$m^2_{\pi 0}=m^2+\frac{\lambda}{6}\Phi^2$,
$\rho^2=\sigma^2+\vec\pi^2$ and 
$\delta L_{ct}[\sigma,\pi^a,\psi]$ is the counterterm Lagrangian. 
For fermions the chemical potential is introduced by the replacement:
$\partial_t\rightarrow \partial_t-i\mu$. One can see that the fermions 
start contributing at level $\mathcal{O}(1/\sqrt{N})$ 
in the large N expansion, 
that is at an intermediate level between LO and NLO of the mesonic sector. 
Since we want to calculate fermionic corrections to the LO which has O(N) 
symmetry we disregard the second independent quartic coupling of the 
$SU(N_f)\times SU(N_f)$ linear $\sigma$ model which is present for $N_f> 2$
and is proportional with the product of two totally symmetric structure
constants $d$. 

Our method consists in taking into account perturbatively 
$N_f=\sqrt{N+1}$ flavors of $N_c$ coloured quarks to 
${\mathcal O}(1/\sqrt{N})$ at one loop order
both in the EoS
\be
V'_\textrm{eff}(\Phi)/\sqrt{N}=:\Phi H(\Phi^2)=
\Phi\left[m^2+\frac{\lambda}{6}\Phi^2+
\frac{\lambda}{6}T_B(M)-\frac{4g^2N_c}{\sqrt{N}}T_F(m_q)\right]=0,
\label{Eq:EoS}
\ee  
and in the gap equation $iG_\pi^{-1}(p^2=M^2)=0$ 
which resums the superdaisy diagrams made of
pions with one possible one loop fermion insertion. 
$T_{B(F)}$ is the bosonic (fermionic) tadpole contribution with mass
$M(m_q)$. The inverse pion propagator reads as
\be
iG_\pi^{-1}(p)=p^2-m^2-\frac{\lambda}{6}\Phi^2
-\frac{\lambda}{6}T_B(M)-B_{1\psi}(p).
\label{Eq:pion}
\ee
The contribution of the fermionic one-loop bubble (``fish'' diagram) 
$B_{1\psi}$ can be related to the fermionic tadpole:
\be
B_{1\psi}(p)=-\frac{4g^2N_c}{\sqrt{N}}T_F(m_q)
+\frac{2g^2N_c}{\sqrt{N}}p^2 I_F(p,m_q),
\label{Eq:bubble-tadpole}
\ee
where $I_F(p,m_q)$ is the scalar bubble integral with external
momentum $p$ evaluated with mass $m_q$ and with 
Fermi-Dirac distribution, $f_F^\pm(\omega)=1/(\exp(\beta(\omega\mp\mu))+1)$.
Substituting relation (\ref{Eq:bubble-tadpole})
into (\ref{Eq:pion}) and making use of
(\ref{Eq:EoS}) one finds that for $p^2=0$ the inverse propagator vanishes,
that is Goldstone's theorem is fulfilled at the minimum of the effective
potential i.e. $M=0$. Another way for seeing this is to use the definition
of $H(\Phi^2)$ given in (\ref{Eq:EoS}) in the expression of the gap
equation. Then one has
\begin{equation}
H(\Phi^2) = M^2\left(1- \frac{2g^2N_c}{\sqrt{N}} I_F(M,m_q)\right),
\label{Eq:M^2}
\end{equation}
which shows that in the spontaneously symmetry broken regime defined by
$H(\Phi^2)=0$, $\Phi\ne0$ the physical pion mass is always zero.

\section{Analytical determination of the TCP}\label{sec:analysys}

\begin{figure}
\includegraphics*[width=6.5cm]{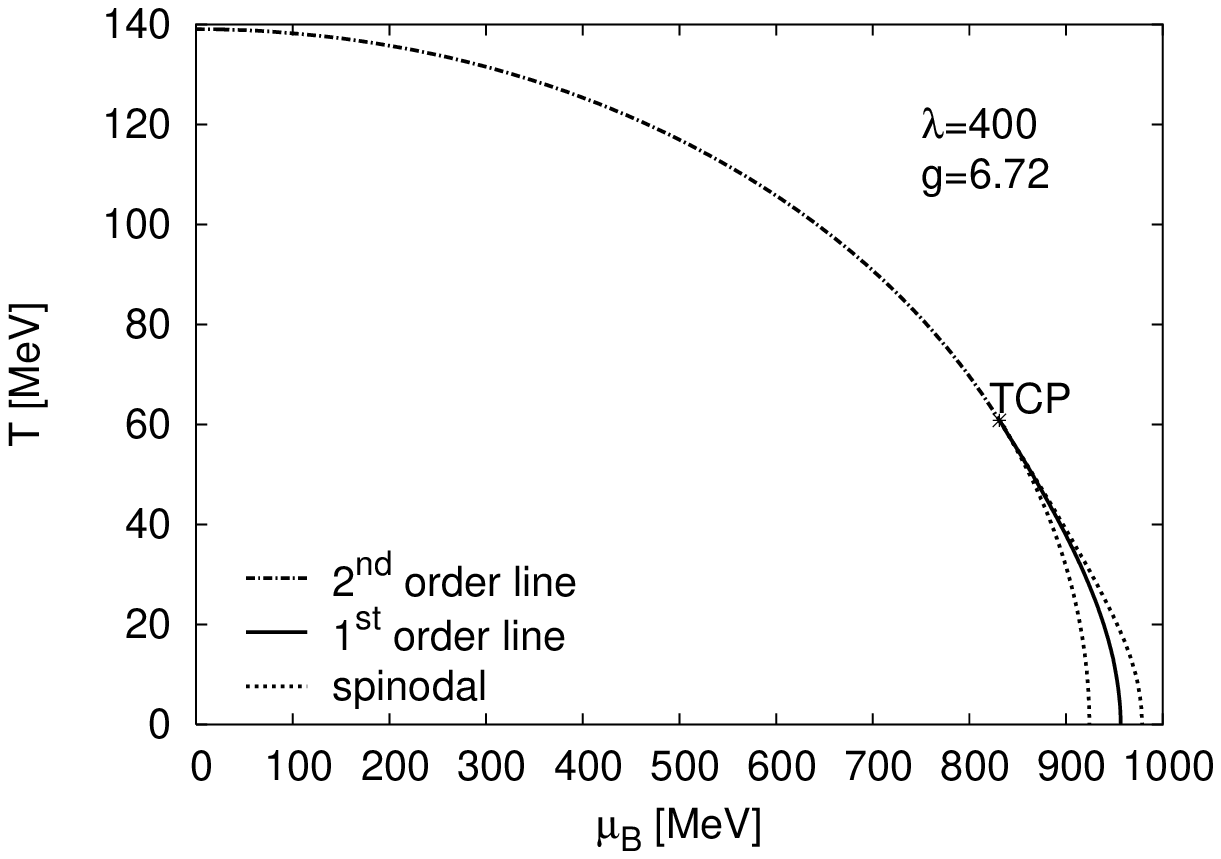}
\hspace*{-0.3cm}
\includegraphics*[width=6.5cm]{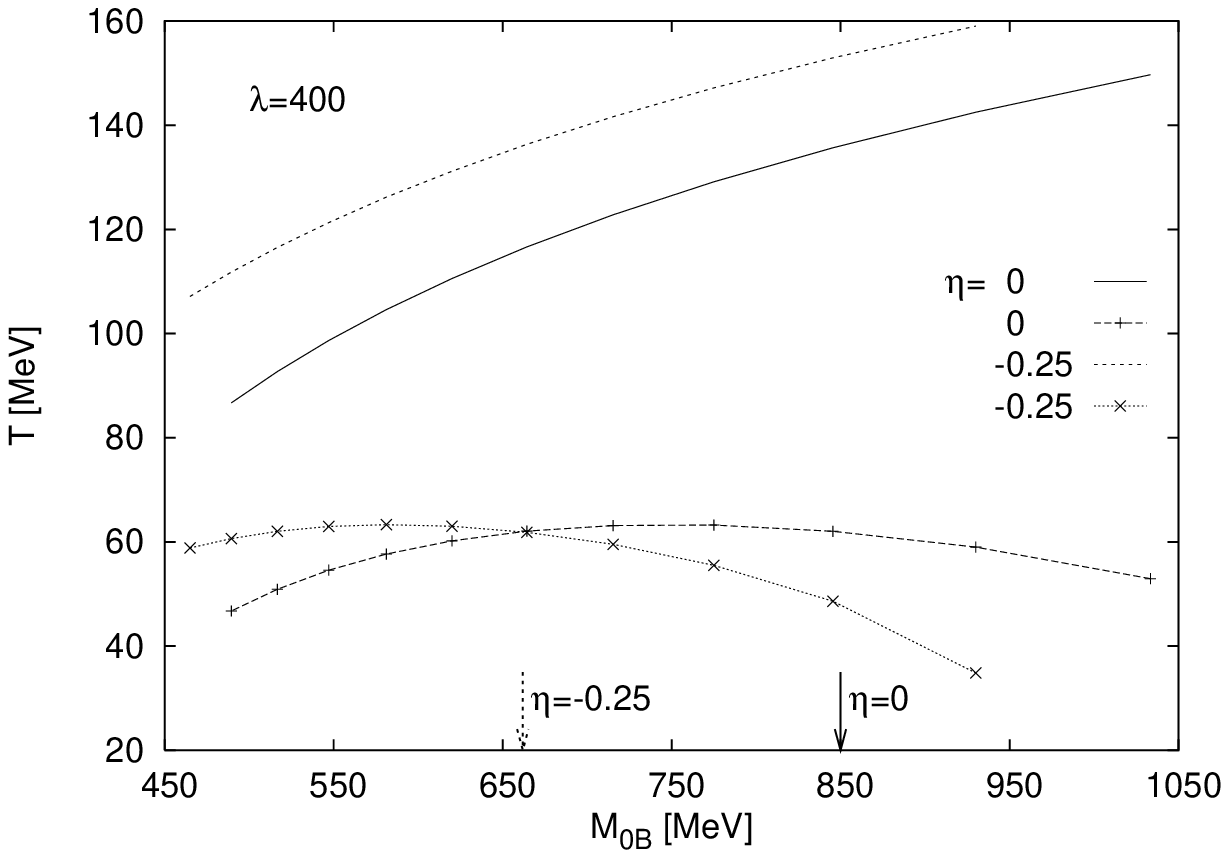}
\vspace*{-0.9cm}
\caption[]{L.h.s.: The $T-\mu$ phase diagram for $M_{0B}=886$ MeV and
$\eta=0$. R.h.s.  the dependence of $T_c(\mu =0)$ (upper curves) and of
$T_{TCP}$ (lower curves) on $M_{0B}$ for $\eta=0$ and $-0.25$
(for the definitions see the text).}
\label{Fig}
\end{figure}

One can choose to study the phase structure by expanding in powers 
of $\Phi$ either the EoS or the effective potential and investigate
its behaviour near the transition.
In ref. \cite{QMM03} we chose the first possibility and determined the
line of continuous transitions by requiring the vanishing of the square
bracket of (\ref{Eq:EoS}) at $\Phi=0$. The pion mass was fixed at $M=0$ 
and only the fermion tadpole was expanded. The line of the 
2$^{\textrm{nd}}$ order phase transitions comes from the condition
\be
m_T^2:=m^2+\frac{g^2\mu^2}{4\pi^2}N_c+
\left(\frac{\lambda}{72}+\frac{g^2N_c}{12}\right)T^2=0.
\label{Eq:quadrat}
\ee
The location of the TCP is determined by requiring
in addition to $m_T^2=0$ also
the vanishing of the coefficient of $\Phi^2$ in the square
bracket of (\ref{Eq:EoS})
\be
\frac{\bar\lambda}{6}:=
\frac{\lambda}{6}+\frac{g^4N_c}{4\pi^2}\left[\frac{\partial}{\partial n}
\Big( Li_n(-e^{-\mu/T})+Li_n(-e^{-\mu/T})
\Big)\Big|_{n=0}- \ln\frac{c_1T}{M_{0B}}\right]=0.
\label{Eq:quartic}
\ee
Here $\ln (c_1/2)=1-\gamma_E+\eta$ and $\eta=\ln(M_{0B}/M_{0F})$ gives a
relation
between $M_{0B}$ and $M_{0F}$, the renormalisation scales of the bosonic and 
fermionic tadpoles. The parameters of the model are fixed at $T=\mu=0$. 
Choosing $m_{q0}=m_N/3\approx 312.67$ and requiring $\Phi_0=f_\pi/2$ for the 
solution of the EoS in the case of two flavors ($N_f=2$) we get $g=6.72$. $\lambda=400$ is fixed by requiring a 
good agreement between the location of the complex pole of the sigma 
propagator on the 2$^{\textrm{nd}}$ Riemann sheet and the experimentally 
favoured mass and width of the sigma particle. Then one can solve the above
equations and find the line of 2$^{\textrm{nd}}$ order phase transitions, 
the first
spinodal line and the location of TCP. This is shown in the l.h.s. of 
Fig.~\ref{Fig}. The other lines in the figure are obtained numerically. In
the r.h.s.  we see the variation of the TCP and critical temperature with
$\eta$ and $M_{0B}$. The study of the pole structure of the $\sigma$
propagator restricts the allowed range of $M_{0B}$ to the right from the
arrow, appearing on the horizontal axis of the figure. In this region 
no low scale
imaginary poles appear besides the well known large scale tachyon, whose
scale increases with decreasing values of the coupling constant $\lambda$. 
In this region one can arrange very easily $T_c(\mu=0)$ to occur in the
region 150-170 MeV but the TCP stays robustly below $70$ MeV. Previous 
effective model studies \cite{LEeff}, and a recent renormalisation group
improved investigation \cite{Wambach}, give similar values which are 
very low when compared with the CEP temperature of the QCD obtained in
\cite{katz04} for physical number of quark flavors and masses. 
This difference could mean that 
the phase transition is not driven by the light degrees of freedom but
by higher excited hadronic states. This possibility was supported by 
\cite{Redlich} when trying to reproduce lattice results on
quark number susceptibilities and pressure with a non-interacting hadron
resonance gas model. It was shown in \cite{Kaj} using 
renormalisation group techniques that the inclusion of gluon degrees of 
freedom is important around $T=150$ MeV. On the other hand we should 
keep in mind that we worked in the limit of an infinitely 
heavy strange quark, and that lattice studies revealed a very strong 
dependence on the values of quark masses \cite{Philippe}.

We now discuss a second method, in which we keep  the $\Phi$-dependence of
the mass gap $M$. The effective potential is obtained by integrating the
functional form of EoS away from equilibrium. This analysis reveals an
interesting property of the $O(N)$ models in the large N limit: the
vanishing of the effective coupling at the critical point \cite{Wetterich}. 
To see this let us first note that $\Phi=0$ becomes a minimum of the
potential as $T\to T_c$ and so $M\to 0$. The fourth derivative of the
effective potential at this point is $V_\textrm{eff}''''(\Phi=0)=4H'(0)$
where
$H'(\Phi^2)=\frac{d M^2}{d \Phi^2}\big(1-\frac{2g^2 N_c}{\sqrt{N}}
I_F(M,m_q)\big)-\frac{2g^2N_c}{\sqrt{N}}M^2\frac{I_F(M,m_q)}{d\Phi^2}$.
Since the last term vanishes at $T_c$ one has to study the behaviour of
$\frac{d M^2}{d \Phi^2}$ which can be formally expressed from the derivative 
of EoS with respect to $\Phi^2$: 
$$
\frac{d M^2}{d\Phi^2}=
\frac{\frac{\lambda}{6}-\frac{4g^4N_c}{\sqrt{N}}I_{F}(0,m_q)
+\frac{2g^2N_c}{\sqrt{N}}M^2\frac{d
I_{F}(M,m_q)}{d\Phi^2}}
{1-\frac{\lambda}{6}
I_{B}(0,M)-\frac{2g^2N_c}{\sqrt{N}}I_{F}(M,m_q)}.
$$
Using a high temperature expansion 
$T_B(M^2)\simeq\frac{T^2}{12} - \frac{MT}{4\pi} + \frac{M^2}{8\pi^2}
  \ln\frac{c_2T}{M_{0B}}$ 
and one has
$
\frac{d T_B(M)}{d M^2} \stackrel{M\to0}{\longrightarrow} -\frac{T}{8\pi M},
$
which gives $\frac{d M^2}{d \Phi^2}\big|_{\Phi=0}=\frac{48\pi M}{\lambda}
(\frac{\lambda}{6}-\frac{4g^4 N_c}{\sqrt{N}}\frac{d T_F(m_q)}{d m_q})
\stackrel{M\to0}{\longrightarrow}0$. 

This means that along the line of $2^{\textrm{nd}}$ order phase transitions
$V_\textrm{eff}''''$ vanishes. Still one should be able to provide a link
between the effective potential of the model and Landau's theory of the
tricritical point in which the location of the TCP is obtained from the
condition of a vanishing quartic coupling. In order to achieve this, we first
perform a high temperature expansion in $H(\Phi^2)$ given by (\ref{Eq:EoS})
which yields
\begin{equation}
H(\Phi^2) = m_T^2 +\frac{\bar\lambda}6\Phi^2 + \kappa\Phi^4 - 2UM +
  \frac{\lambda M^2}{48\pi^2}\ln\frac{c_2T}{M_0},
\end{equation}
where 
$U=\frac{\lambda T}{48\pi}$ and 
$\kappa=\frac{g^6N_c}{16\pi^2\sqrt{N}}\frac{1}{T^2}\frac{\partial}{\partial n}
\left[\mathrm{Li}_n(-e^{\mu/T})+ \mathrm{Li}_n(-e^{-\mu/T})
\right]\big|_{n=-2}$. In the following discussion for the sake of
simplicity we omit in (\ref{Eq:M^2}) the one-loop fermion 
contribution $I_F(M,m_q)$. We then rewrite the resulting $H(\Phi^2)=M^2$
relation as an equation which determines the $\Phi^2$ dependence of $M$
\be
 W M^2 + 2UM -
\left(m_T^2 +\frac{\bar\lambda}{6}\Phi^2 +\kappa \Phi^4\right)=0.
\label{Eq:Mapprox}
\ee
where $W=1-\frac{\lambda^2}{48\pi^2}\ln\frac{c_2T}{M_0}$ and
$\ln \frac{c_2}{4\pi} = \frac{1}{2}-\gamma$. 
The solution of (\ref{Eq:Mapprox}) approximates $H(\Phi^2)$, so after
substituting it into (\ref{Eq:EoS}) it can be integrated to obtain an 
approximation to the effective potential. Subsequent expansion in
powers of $\Phi^2$ yields
\be
V_\mathrm{approx}(\Phi)/\sqrt{N} = \frac{m_T^2}{4U^2W}\left[
\frac{m_T^2}2\Phi^2 +\frac{\bar\lambda}{12} \Phi^4+
\left(\frac{W\bar\lambda^2}{216m_T^2}+\frac{\kappa}{3}\right)\Phi^6
\right].
\label{eq:Veff_Tc}
\ee
A few remarks are in order here. {\it First of all} whether this potential 
has the right shape, and whether this approximation has the correct physical
meaning depends on the actual values of the parameters. One can easily
verify that for the parameters used in drawing Fig.~\ref{Fig}, $W<0$ in the
temperature range of the $2^{\textrm{nd}}$ order phase transition.
For increasing values of $\mu/T$, $\kappa$ changes sign at 1.91
from negative to positive values. Since around TCP $\kappa$
has to be positive, this sign change restricts the location of
the TCP to large values of $\mu$: it must be in the region $\mu>2T$.
{\it Second}, in the broken symmetry phase $m_T^2<0$ and it vanishes at $T_c$
meaning that the coefficient of the quartic term in $\Phi$ vanishes also. 
Around $T_c$ from (\ref{Eq:quadrat}) one gets $m_T^2\sim T-T_c$. Therefore 
the scaling of the effective mass 
$m_\mathrm{eff}=\big|\frac{m_{T}^2}{2U W}\big|\sim|T-T_c|$
gives the correct leading order critical exponent $\nu=1$ for the O(N)
model  at large $N$. {\it Third}, to the expression in the square bracket 
in (\ref{eq:Veff_Tc}) the usual Landau type analysis applies. 
The location of the TCP is determined by the same conditions as in the
analysis based on EoS, namely (\ref{Eq:quadrat}) and 
(\ref{Eq:quartic}). To obtain the scaling exponent of
the order parameter on the second order line we need only the first two
terms in (\ref{eq:Veff_Tc}). The minimum condition gives
$\Phi^2=-\frac{3m_T^2}{\bar\lambda}\,\Rightarrow
\Phi\sim(T-T_c)^{\beta}$, $\beta=1/2$. At TCP one can 
set $\bar\lambda=0$ in (\ref{eq:Veff_Tc}) and keep the sixth order term. The
minimum condition gives
$\Phi^4=-\frac{m_T^2}{2\kappa}\,\Rightarrow \Phi\sim(T-T_c)^{\beta}$, 
$\beta=1/4$.
We verified numerically using the exact EoS that the value of $\beta$ on the
second order line and at the TCP agrees within the numerical accuracy with
their mean field estimates given above.

Refinement of the approximation is needed if one wants to enter the
metastable region between the spinodals where $m_T^2>0$ and $\bar\lambda<0$, 
but we think that nevertheless the presented analysis captures 
essential features of the phenomenon.

\section{Conclusions}\label{sec:concl}
In the $1/\sqrt{N}$ order of the large N approximation to the solution
of the chiral quark-meson
model we presented in the chiral limit an analytical determination of 
the phase diagram and of the location of the tricritical point. We have
shown that with a reasonable approximation which works fine at high
temperature, the effective potential of the theory can be related to 
mean field Landau type potential, from which the critical exponents can be
determined.

In order to assess to what extent a description of the phase diagram of QCD is
possible using a low-energy effective model, one has both to improve
our method of solving the model and to get as close as possible to the
physical case of $2+1$ flavors. The first issue necessitates taking into
account the full momentum dependence of the self-consistent pion propagator,
the implementation of a new resummation method and solving the
Dyson-Schwinger equation for the fermions. The latter problem requires the
extension of our method to the $SU(3)_L\times SU(3)_R$ case.

\section*{Acknowledgement(s)}
Zs. Sz. thanks the Institute for Nuclear Theory at the University of
Washington for its hospitality and the Department of Energy for partial
support during the completion of this work.
 The authors acknowledge the support of the Hungarian Research Fund (OTKA)
under contract numbers F043465, T034980, and T037689. 

\section*{Notes}
\begin{notes}
\item[a]
E-mail: jakovac@planck.phy.bme.hu
\item[b]
E-mail: patkos@ludens.elte.hu
\item[c]
Speaker, E-mail: szepzs@antonius.elte.hu
\item[d]
E-mail: psz@galahad.elte.hu
\end{notes}

\vfill\eject
\end{document}